\documentclass[epj]{webofc}
\usepackage[varg]{txfonts}   
\renewcommand{\vec}[1]{\mbox{\boldmath $#1$}}
\woctitle{CNR*15}
\begin{document}
\title{Recent developments in heavy-ion fusion reactions around 
the Coulomb barrier}
%
%

\author{K. Hagino\inst{1,2,3}, 
        N. Rowley \inst{4}, \and J.M. Yao\inst{1,5,6}
}

\institute{Department of Physics, Tohoku University, Sendai 980-8578, Japan
\and
Research Center for Electron Photon Science, Tohoku University, 1-2-1 Mikamine, Sendai 982-0826, Japan
\and
National Astronomical Observatory of Japan, 2-21-1 Osawa,
Mitaka, Tokyo 181-8588, Japan
\and
Institut de Physique Nucl\'{e}aire, UMR 8608, CNRS-IN2P3 et Universit\'{e}
de Paris Sud, 91406 Orsay Cedex, France
\and
School of Physical Science and Technology, Southwest University, Chongqing 400715, China
\and
Department of Physics and Astronomy, University of North Carolina,
Chapel Hill, North Carolina 27516-3255, USA 
          }

\abstract{%
The nuclear fusion is a reaction to form a compound nucleus. It 
plays an important role in several circumstances in nuclear physics 
as well as in nuclear astrophysics, such as synthesis of superheavy 
elements and nucleosynthesis in stars. Here we discuss two recent 
theoretical developments in heavy-ion fusion reactions at energies 
around the Coulomb barrier. The first topic is a generalization of 
the Wong formula for fusion cross sections in a single-channel 
problem. By introducing an energy 
dependence to the barrier parameters, we show that the generalized formula 
leads to results practically indistinguishable from a full 
quantal calculation, even for light symmetric systems such as 
$^{12}$C+$^{12}$C, for which fusion cross sections show an 
oscillatory behavior. We then discuss a semi-microscopic modeling
of heavy-ion fusion reactions, which combine 
the coupled-channels approach to 
the state-of-the-art nuclear 
structure calculations for low-lying collective motions. 
We apply this method to 
subbarrier fusion reactions of $^{58}$Ni+$^{58}$Ni 
and $^{40}$Ca+$^{58}$Ni systems, and discuss the role 
of anharmonicity of the low-lying vibrational motions. 
}

\maketitle
\section{Introduction}

Fusion is defined as a reaction in which two separate nuclei combine 
together to form a (hot) compound nucleus. 
It is indispensable to understand its reaction dynamics in order to 
understand synthesis of superheavy elements and nucleosynthesis in stars. 
Moreover, fusion reactions inherently contain rich physics, 
as there is a strong interplay in it, 
especially at energies 
around the Coulomb barrier, 
between the nuclear reaction and the nuclear structure. 
That is, 
it has been known well that 
fusion cross sections at subbarrier energies 
are largely enhanced 
relative to a 
prediction of a one-dimensional potential model, due to the couplings 
of the relative motion between the
colliding nuclei to several nuclear intrinsic motions 
\cite{HT12,DHRS98,BT98,Back14,Canto15}.
This is in marked contrast to high energy nuclear reactions, in which 
the reaction dynamics is much simpler and the distorted wave Born 
approximation (DWBA) often suffices its treatment. 
Fusion also offers a unique opportunity to study quantum tunneling 
with many degrees of freedom. 
That is, fusion takes place only 
by quantum tunneling at energies below the 
Coulomb barrier, and the subbarrier enhancement of fusion cross sections 
can be viewed as a consequence of a coupling assisted tunneling. 
Heavy-ion fusion reactions are unique in this respect because 
a variety of intrinsic degrees of freedom are involved, such as 
a surface vibration with several multipolarities, various sort of nuclear 
deformations and the associated rotational motion, and 
several types of particle transfer processes. 
This is in contrast to atomic and molecular collisions, in which only a 
limited types of intrinsic motion are involved. 
Also, in heavy-ion fusion reactions, the incident energy can be varied 
in order to study the energy dependence of the tunneling probability, 
whereas the energy is fixed in many other tunneling phenomena in 
nuclear physics, such as alpha decays. 

In this contribution, we shall 
discuss two recent theoretical developments
in heavy-ion fusion reactions. 
The first topic is a generalization of the Wong formula
for fusion cross sections \cite{RH15}.
Although this celebrated
formula yields almost exact results for
single-channel calculations for relatively heavy systems
such as $^{16}$O+$^{144}$Sm,
it tends to overestimate the cross section for light systems such
as $^{12}$C+$^{12}$C.
We generalize the formula to take account of the
energy dependence of the barrier parameters and show that the
energy-dependent version
reproduces almost perfectly 
results of a full quantal calculation.
We also discuss the deviations arising from the discrete nature
of the intervening angular momenta, whose effect can lead to
an oscillatory contribution
to the excitation function.

The second topic is a semi-microscopic modeling
of heavy-ion fusion reactions \cite{HY15}.
For this purpose, 
we first describe microscopically 
low-lying collective excitations of atomic nuclei with
the multi-reference covariant density functional theory,
and combine them with coupled-channels calculations.
We use the calculated transition
strengths among several collective states as inputs to the coupled-channels
calculations.
This approach provides a natural way to describe anharmonic multi-phonon
excitations as well as a deviation of rotational excitations
from a simple rigid rotor.
We apply this method to subbarrier fusion reactions of $^{58}$Ni+$^{58}$Ni 
and $^{40}$Ca+$^{58}$Ni systems, and discuss the role of anharmonicity 
in subbarrier fusion of these systems. 

\section{Generalization of the Wong formula for fusion cross sections}

Let us first discuss the well-known Wong formula for fusion cross sections. 
The simplest approach to heavy-ion fusion reactions is 
to use the potential model, assuming that 
both the projectile and the target 
nuclei are inert.  
Fusion cross sections are then obtained by calculating 
the $S$-matrix, $S_l$, 
for each partial wave. 
If all the flux crossing the Coulomb barrier fuses, 
the fusion cross sections $\sigma_{\rm fus}$ read
\begin{equation}
\sigma_{\rm fus}(E)=\frac{\pi}{k^2}\sum_{l=0}^\infty(2l+1)P_l(E),
\label{fus}
\end{equation}
where $k=\sqrt{2\mu E/\hbar^2}$ is the wave number associated with the energy 
$E$ ($\mu$ being the reduced mass) 
and $P_l(E)=1-|S_l|^2$ is the penetrability of the Coulomb barrier 
for the $l$-th 
partial wave. 

Wong has derived a compact approximate expression for 
Eq. (\ref{fus}) \cite{Wong73}, which has by now 
been known as the Wong formula. 
To this end, Wong introduced the following three approximations: 
\begin{itemize}
\item {\it the parabolic approximation.}  
The total (that is, the nuclear + the Coulomb) 
inter-nucleus potential $V_0(r)$ for the $s$-wave 
is approximated by a 
parabolic function, that is, 
\begin{equation}
V_0(r)\sim V_b -\frac{1}{2}\mu \Omega^2 (r-R_b)^2,
\end{equation}
where $V_b$, $R_b$, and $\hbar\Omega$ are the barrier height, the barrier 
position, and the ``barrier curvature'', respectively. 

\item 
{\it the $l$-independent barrier position and curvature.} 
The barrier position, $R_b$, and the barrier curvature, $\hbar\Omega$, 
are assumed to be independent of $l$. 
In this case, the effective potential for the $l$-th partial wave 
reads, 
\begin{equation}
V_0(r)+\frac{l(l+1)\hbar^2}{2\mu r^2} \sim 
V_b +\frac{l(l+1)\hbar^2}{2\mu R_b^2} 
-\frac{1}{2}\mu \Omega^2 (r-R_b)^2. 
\end{equation}
The penetrability can then be calculated analytically 
with the Hill-Wheeler formula \cite{HW53} as, 
\begin{equation}
P_l(E)=\frac{1}{1+\exp\left[\frac{2\pi}{\hbar\Omega}\left(V_b
+\frac{l(l+1)\hbar^2}{2\mu R_b^2} -E\right)\right]}.
\end{equation}

\item {\it the continuum approximation for $l$.} 
The angular momentum $l$ is treated as a continuous variable and the sum in 
Eq. (\ref{fus}) is replaced by the integral, 
\begin{equation}
\sigma_{\rm fus}(E)=\frac{\pi}{k^2}\sum_{l=0}^\infty(2l+1)P_l(E)
\to 
\frac{\pi}{k^2}\int^\infty_0dl\,(2l+1)P_l(E).
\label{eq:continuum}
\end{equation}

\end{itemize}
With these approximations, the fusion cross sections are 
obtained as \cite{Wong73,HT12,BT98},
\begin{equation}
\sigma_{\rm fus}(E)=\frac{\hbar\Omega}{2E}R_b^2 \ln\left[1+\exp\left(
\frac{2\pi}{\hbar\Omega}(E-V_b)\right)\right].  
\label{wong}
\end{equation}
Notice that this formula yields the classical fusion cross sections, 
\begin{equation}
\sigma_{\rm fus}(E)\sim
\pi R_b^2 \left(1-\frac{V_b}{E}\right), 
\end{equation}
at energies well above the Coulomb barrier, $E-V_b \gg \hbar\Omega/2\pi$. 

\begin{figure}
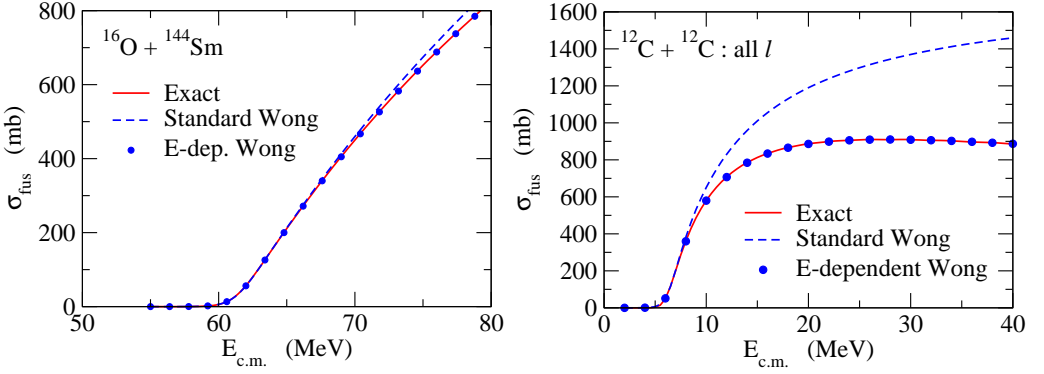

\centering
\includegraphics[scale=0.45,clip]{fig1a}
\includegraphics[scale=0.45,clip]{fig1b}
\caption{
Fusion cross sections for the 
$^{16}$O+$^{144}$Sm (the left panel) and the
$^{12}$C+$^{12}$C (the right panel) systems. 
The solid lines show the results of a full quantal 
calculation, while the dashed line shows the fusion cross 
sections evaluated with the Wong formula, Eq. (\ref{wong}). 
The filled circles, on the other hand, denote the results of the generalized 
(the energy dependent) Wong formula given by Eq. (\ref{wong2}). 
The Bose symmetry of the identical spin-0 system is ignored 
for the 
$^{12}$C+$^{12}$C system, and both even and odd partial waves are summed up 
in the cross sections. 
}
\end{figure}

The left panel of Fig. 1 shows the performance of the Wong formula 
for the $^{16}$O+$^{144}$Sm system. The solid line is obtained quantum 
mechanically by solving the Schr\"odinger equation for a given inter-nucleus 
potential, while the dashed line is obtained with the Wong 
formula, Eq. (\ref{wong}). 
Even though the Wong formula slightly overestimates the 
fusion cross sections at energies well 
above the Coulomb barrier, reflecting 
the fact that the barrier position gradually decreases as a function 
of $l$, the overall agreement is satisfactory. 
In contrast, the situation drastically changes for light systems. 
The right panel of Fig. 1 shows a comparison between the quantal and 
the approximate fusion cross sections for the 
$^{12}$C+$^{12}$C system. Here, in order to simplify the discussion, 
for the moment we ignore the Bose symmetry of this identical 
spin-0 system and sum over all even and odd partial waves. 
One can see that the Wong formula largely overestimates fusion 
cross sections, and the agreement is much worse than the heavier system, 
$^{16}$O+$^{144}$Sm. This is because the centrifugal potential 
plays a more important role in light systems, and the Coulomb barrier 
is not rigid against a variation of angular momentum \cite{RH15}. 

One can improve the Wong formula by introducing 
the energy dependence to the barrier 
parameters $[V_b,R_b,\hbar\Omega]$. 
To this end, we first introduce the grazing angular momentum $l_g$, 
at which the barrier height of the effective potential becomes identical 
to the incident energy, $E$. If we denote the barrier position for $l_g$ 
as $R_E$, $l_g$ and $R_E$ are related to each other as, 
\begin{equation}
\frac{l_g(l_g+1)\hbar^2}{2\mu R_E^2}=E - V_E,
\end{equation}
where $V_E=V_0(R_E)$ is the sum of the Coulomb and the nuclear 
potentials at $R_E$. 
An idea of the generalized Wong formula is to replace the barrier parameters 
$[V_b,R_b,\hbar\Omega]$ with the energy dependent ones, 
$[V_E,R_E,\hbar\Omega_E]$, where $\hbar\Omega_E$ 
is the barrier curvature for the grazing angular momentum $l_g$. 
That is, the generalized version of the 
Wong formula now reads \cite{RH15}, 
\begin{equation}
\sigma_{\rm fus}(E)=\frac{\hbar\Omega_E}{2E}R_E^2 \ln\left[1+\exp\left(
\frac{2\pi}{\hbar\Omega_E}(E-V_E)\right)\right]. 
\label{wong2}
\end{equation}
Notice that we still assume an $l$-independent barrier position 
and curvature in 
integrating Eq. (\ref{eq:continuum}). However, 
the generalization 
is such that these parameters are evaluated 
at the grazing angular 
momentum for each energy $E$, rather than using the values for the $s$-wave. 
The filled circles in Fig. 1 show the fusion cross sections obtained with the 
generalized Wong formula, Eq. (\ref{wong2}). 
One can see that the quantal results are well reproduced 
by introducing the energy dependence to the barrier parameters, both 
for the $^{16}$O+$^{144}$Sm and the $^{12}$C+$^{12}$C systems. 

\begin{figure}
\centering
\includegraphics[scale=0.45,clip]{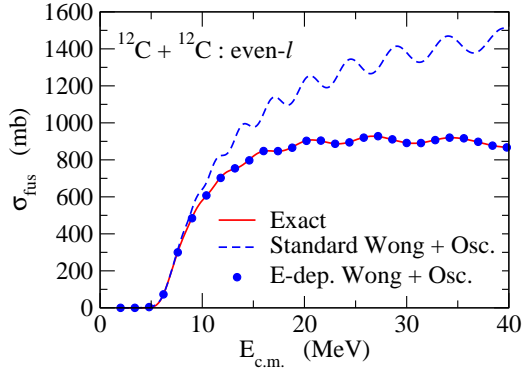}
\caption{
Fusion cross sections for the $^{12}$C+$^{12}$C system obtained with even partial waves only. 
The solid line shows the exact quantal result. 
The filled circles and the dashed line denote 
the results of the Wong 
formula supplemented with the oscillatory contribution, Eq. (\ref{osc}), 
with and without the energy dependence of the barrier parameters, 
respectively. 
}
\end{figure}

One can also go beyond the continuum approximation for the angular 
momentum sum. Using the Poisson sum formula \cite{Brink}, 
the next order correction to the Wong formula reads \cite{RH15,Poffe}, 
\begin{equation}
\sigma_{\rm fus}(E)\sim \sigma_{\rm Wong}+\sigma_{\rm osc},
\end{equation}
where 
$\sigma_{\rm Wong}$ is the smooth part of 
fusion cross sections given by the Wong formula, Eq. (\ref{wong2}), 
while the oscillatory contribution 
$\sigma_{\rm osc}$ is given by \cite{RH15,Poffe},
\begin{equation}
\sigma_{\rm osc}=2\pi R_E^2\frac{\hbar\Omega_E}{E}e^{-2\xi}\sin(2\pi l_g),
\end{equation}
with 
\begin{equation}
\xi=\frac{\hbar\Omega_E}{2l_g+1}\cdot\frac{\pi\mu R_E^2}{\hbar^2}.
\end{equation}
Usually, the oscillatory contribution, $\sigma_{\rm osc}$, is much smaller than the 
smooth part given by $\sigma_{\rm Wong}$. There are, however, certain circumstances 
where the oscillatory contribution becomes significantly large. 
This is for a light symmetric 
system of two identical spin-0 bosons, in which all odd partial waves 
disappear due to the symmetrization of the wave function. In this case, 
the fusion cross section is 
given by twice the sum over the even partial waves. 
The resultant formula now becomes \cite{RH15,Poffe},
\begin{equation}
\sigma_{\rm osc}=2\pi R_E^2\frac{\hbar\Omega_E}{E}e^{-\xi}\sin(\pi l_g),
\label{osc}
\end{equation}
whereas the smooth part is still given by the Wong formula, Eq. (\ref{wong2}). 
The oscillatory part is now significantly larger than before, 
since the negative exponent has been reduced by a factor of 2, that is 
$e^{-2\xi}\to e^{-\xi}$. 

Figure 2 shows fusion cross sections for the $^{12}$C+$^{12}$C system, obtained by 
taking only the even partial waves. The meaning of each line is the same as in Fig. 1. 
One can see that the fusion cross sections significantly 
oscillate as a function of $E$. 
The generalized Wong formula, supplemented  
with the oscillatory correction, gives an excellent approximation to the exact results. 

\section{Semi-microscopic modeling of heavy-ion fusion reactions with 
a beyond-mean field method} 

The potential model presented in the previous section works for light 
systems, such as $^{14}$N+$^{12}$C, 
but 
it largely underestimates fusion cross sections for 
heavier systems, such as $^{16}$O+$^{154}$Sm, at energies below the 
Coulomb barrier \cite{HT12}. 
It has been well recognized by now that this large enhancement of 
subbarrier fusion cross sections is caused by the couplings of the 
relative motion between the colliding nuclei to several nuclear intrinsic 
degrees of freedom, such as low-lying collective excitations in the 
colliding nuclei as well as several nucleon 
transfer processes \cite{HT12,DHRS98,BT98,Back14}. 
A natural framework for heavy-ion subbarrier fusion has thus been 
the coupled-channels method 
with relevant degrees of freedom \cite{HT12,HRK99}. 
This approach has not only successfully accounted for 
the subbarrier enhancement 
of fusion cross sections for many systems but 
has also provided a natural interpretation of the so called fusion 
barrier distributions \cite{HT12,DHRS98,RSS91,LDH95}. 

In the coupled-channels approach,
if the projectile nucleus is assumed to be inert, 
one expands the total wave function of the system 
in terms of the eigen-functions
of the collective states in the target nucleus, $|\varphi_{I0}\rangle$,
as
\begin{equation}
  \Psi_{LM_L}(\vec{r})=\sum_I\frac{u_I(r)}{r}Y_{LM_L}(\hat{\vec{r}})
  |\varphi_{I0}\rangle,
\label{totwf}
\end{equation}
where 
$I$ and $L$ are the angular momenta for the target state 
and for the relative motion, respectively.
Here, we have 
introduced the isocentrifugal
approximation \cite{HT12}, 
and have assumed that $L$ and $M_L$ (thus also $M_I$) do not change 
by the excitation of the target nucleus. 
Substituting Eq. (\ref{totwf}) to the projected Schr\"odinger
equation, $\langle \varphi_{I0}|H-E|\Psi_{LM_L}\rangle=0$, 
the coupled-channels equations for the radial wave functions
$u_I(r)$ read \cite{HT12},
\begin{equation}
\left[-\frac{\hbar^2}{2\mu}\frac{d^2}{dr^2}
    +\frac{L(L+1)\hbar^2}{2\mu r^2}
    +V_0(r)
-E+\epsilon_I\right]u_I(r) 
+\sum_{I'}V_{II'}(r)u_{I'}(r)=0,
\label{cceq}
\end{equation}
where 
$\epsilon_I$ is the energy of the target state $I$.
$V_{II'}(r)$ are the coupling matrix elements given by
\begin{equation}
  V_{II'}(r)=\langle\varphi_{I0}|V_{\rm coup}(r,\alpha_{\lambda0})|\varphi_{I'0}
  \rangle,
\label{coupV}
\end{equation}
where $V_{\rm coup}$ is the coupling potential and $\alpha_{\lambda 0}$ is
the excitation operator with a multipolarity $\lambda$.

\begin{figure}
\centering
\includegraphics[scale=0.5,clip]{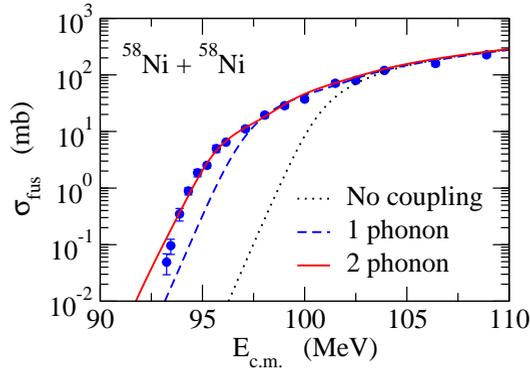}
\caption{The results of coupled-channels calculations for 
the fusion cross sections for the 
$^{58}$Ni+$^{58}$Ni system. 
The vibrational 
coupling to multi-quadrupole-phonon 
states are considered in the harmonic oscillator approximation. 
The experimental data are taken from Ref. \cite{Beckerman81}. 
}
\end{figure}

In heavy-ion fusion reactions at energies around the Coulomb 
barrier, multiple excitations to higher members of collective 
modes, such as multi-phonon states and high-spin states in the 
ground state rotational band, often play an 
important role \cite{DHRS98,LDH95,Stefanini95}. 
As an example, Fig. 3 shows 
fusion cross sections for the 
$^{58}$Ni+$^{58}$Ni 
system obtained with the coupled-channels 
calculations which take into account 
the vibrational coupling 
to quadrupole phonon states. 
One can see that 
the coupling to the first excited state is insufficient 
in order to account 
for the subbarrier enhancement of fusion cross sections for this system. 
In order to take into account the multiple excitations in 
coupled-channels calculations, 
one usually uses the simple harmonic oscillator model for vibrational 
nuclei \cite{HT12}. 
In this approximation, the $n$-phonon multiplet states are combined 
together to form a single effective channel as \cite{HT12}
\begin{equation}
|n\rangle = \frac{1}{\sqrt{n!}}\left(a_{\lambda 0}^\dagger\right)^n|0\rangle,
\end{equation}
where $a_{\lambda 0}^\dagger$ is the phonon creation operator and 
$|0\rangle$ is the vacuum state. 
The energy of this channel is given by $\epsilon_n=n\hbar\omega$, where 
$\hbar\omega$ is the oscillator quantum, and the matrix elements 
of $\alpha_{\lambda 0}$ are related to the phonon numbers as 
$\langle n|\alpha_{\lambda 0}|n'\rangle \propto 
\sqrt{n'}\delta_{n,n'-1}+\sqrt{n'+1}\,\delta_{n,n'+1}$. 
The solid line in Fig. 3 shows the result of the two phonon couplings. 
One can see that the subbarrier fusion enhancement is now well reproduced. 

Even though the harmonic oscillator approximation appears to work well 
for the $^{58}$Ni+$^{58}$Ni system, 
most nuclei, including $^{58}$Ni, do not have a 
pure harmonic oscillator
spectrum in reality. 
Concerning the $^{58}$Ni nucleus, 
the degeneracy of the experimentally known 
two-phonon triplet is considerably broken (see Fig. 4), and also 
a recent theoretical calculation 
indicates that the $B(E2)$ strengths
among the collective levels in $^{58}$Ni 
deviate largely from the harmonic oscillator limit \cite{Yao15}. 
Such anharmonic vibrations can be described, {\it e.g.,} 
a multi-reference density-functional
theory (MR-DFT), which has been rapidly developed 
for the past decade \cite{Bender03,Yao14}.
This method is based on the so called beyond-mean-field 
approximation, which incorporates 
the angular momentum and particle number projections 
for a mean-field wave function. 
The quantum fluctuation of the mean-field wave function 
is also taken into account with 
the generator coordinate method (GCM). 

\begin{figure}
\centering
\includegraphics[scale=0.7,clip]{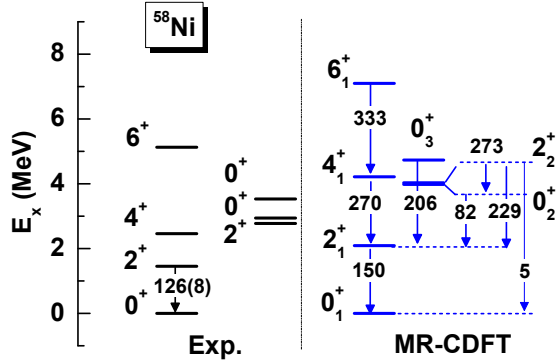}
\caption{
The low-lying energy spectra of $^{58}$Ni 
obtained with the 
multi-reference covariant density functional 
theory (MR-CDFT) method with the PC-PK1 force. 
The arrows indicate the 
$E2$ transition strengths, given in units of $e^2$fm$^4$. 
The experimental data
are taken from Refs. \cite{NNDC,allmond14}. } 
\end{figure}

Figure 4 shows 
the spectrum of the $^{58}$Ni nucleus \cite{HY15} constructed with 
the MR-DFT calculation 
with 
the covariant density functional with 
the PC-PK1 interaction \cite{PC-PK1}. 
One can see that 
the excitation energies of the 0$^+_2$, 2$^+_2$, and 4$^+_1$ states
are about twice the energy of the $2^+_1$ state. 
On the other hand, it is interesting to notice that 
the overall pattern of $B(E2)$ values is quite different
from what would be expected for a harmonic vibrator, 
in which 
the $B(E2)$ value from any of 
the two-phonon triplet states to the 2$^+_1$ state is exactly 
twice the $B(E2)$ 
value from the 2$^+_1$ state to the ground state. 
In particular,
the $E2$ transition from the $0^+_2$ to the $2^+_1$ states 
is much smaller than that
for the $4^+_1\to 2^+_1$ and the $2^+_2\to 2^+_1$ 
transitions. 
Instead, the $0^+_2$ state has a strong transition from the 
$2^+_2$ state, which clearly indicates that the $0^+_2$ state 
is not a member of the two-phonon triplet. 
Compared to the $0^+_2$ state, the $E2$
transition strength 
for the $0^+_3\to 2^+_1$ transition 
is much larger and is comparable to that 
for the $4^+_1\to 2^+_1$ and the $2^+_2\to 2^+_1$ 
transitions. 
This suggests that 
the $0^+_3$ state is a better candidate for a member of 
the two-phonon 
triplets than the $0_2^+$ state, even 
though the excitation energy is a little large. 

\begin{figure}
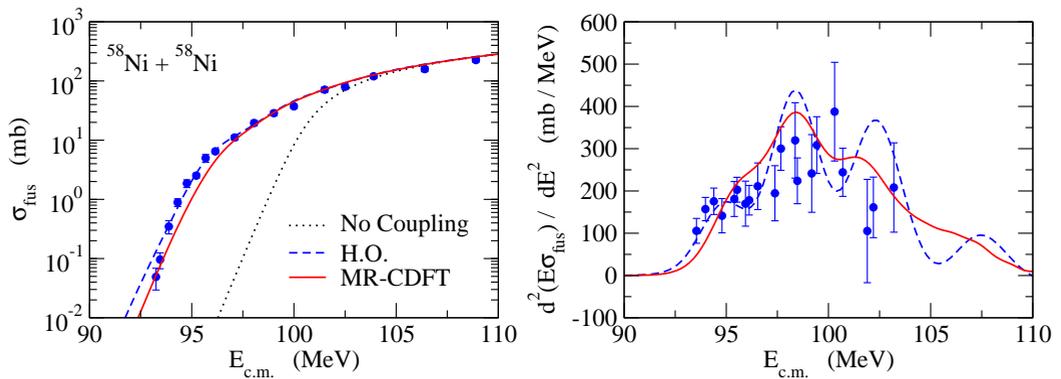

\centering
\includegraphics[scale=0.45,clip]{fig5a}
\includegraphics[scale=0.45,clip]{fig5b}
\caption{
The fusion cross sections (the left panel) and 
the fusion barrier distributions (the right panel) 
for the $^{58}$Ni+$^{58}$Ni system. 
The dashed line is the result of the coupled-channels calculations 
including the double quadrupole phonon excitations in each $^{58}$Ni 
nucleus in the harmonic oscillator limit. 
The solid line, on the other hand, is 
obtained with the multi-reference covariant density functional 
theory (MR-CDFT) method by including 
the couplings to the 
0$^+_1$, 2$^+_1$, 0$^+_3$, 2$^+_2$, and 4$^+_1$ states in $^{58}$Ni. 
The dotted line in the left panel denotes 
the result in the absence of the channel couplings. 
The experimental data are taken from Ref. \cite{Beckerman81} for 
the fusion cross sections and from Ref. \cite{Stefanini95} 
for the fusion barrier distribution. 
}
\end{figure}

In order to take into account such anharmonic nature of the 
vibrational excitations of $^{58}$Ni in the coupled-channels 
calculations, we first replace the operator $\alpha_{\lambda 0}$ 
in Eq. (\ref{coupV}) with the corresponding microscopic multipole 
operator. 
That is, we replace the matrix elements of
$\alpha_{\lambda 0}$ as,  
\begin{equation}
\langle\varphi_{I0}|\alpha_{\lambda0}|\varphi_{I'0}\rangle 
\to 
\frac{4\pi}{3Z_TeR_T^\lambda}
\langle\varphi_{I0}|Q_{\lambda0}|\varphi_{I'0}\rangle,
\label{Vcoup2}
\end{equation}
where 
$Q_{\lambda\mu}=\sum_ir_i^\lambda Y_{\lambda\mu}(\hat{\vec{r}}_i)$ is the 
microscopic multipole operator, 
and $R_T$ and $Z_T$ are the radius and the charge number 
of the target nucleus, respectively. 
$|\varphi_{I'0}\rangle$ on the right hand side 
is a many-body 
wave function evaluated with the MR-CDFT method. 
We still use a phenomenological potential for 
$V_0(r)$ and $V_{\rm coup}(r,\alpha_{\lambda 0})$, and we call our method 
a semi-microscopic approach. 

In heavy-ion fusion reactions, the coupling of the ground state 
to the lowest-lying 
collective state is most important. The strength for such coupling can 
often be estimated from an experimental transition probability. 
Since it is too much to expect that a MR-CDFT calculation agrees perfectly 
with experimental data, 
we introduce an overall scaling factor to the matrix 
elements, Eq. (\ref{Vcoup2}), 
so that the transition from the lowest-lying collective state to the 
ground state is consistent with experimental data. 
The MR-CDFT calculation then provides 
the relative strengths among collective levels, which are often not 
available experimentally. 
The excitation energy, on the other hand, is often known well 
for many levels, and we simply take them in the calculations whenever they 
are available experimentally. 

Figures 5 shows the fusion 
cross section $\sigma_{\rm fus}(E)$ 
and the fusion barrier distribution 
$D_{\rm fus}(E) = d^2(E\sigma_{\rm fus})/dE^2$ \cite{DHRS98,RSS91}
for the $^{58}$Ni+$^{58}$Ni reaction so obtained. 
The dashed line shows the result of the 
coupled-channels calculations including up to the double phonon states  
in the harmonic oscillator limit. 
All the mutual excitations between the projectile and the target nuclei
are included. 
On the other hand, the solid line in the figure is obtained 
with the coupling strengths evaluated with the 
MR-CDFT method. 
To this end, we include 
the 0$^+_1$, 2$^+_1$, 0$^+_3$, 2$^+_2$, and 4$^+_1$ states in $^{58}$Ni 
in the 
coupled-channels calculations.
Again, all the mutual excitation channels are taken into account. 
For comparison, the figure also shows the result of no-coupling limit by 
the dotted line.
One can see that the calculations in the harmonic limit overestimate
fusion cross sections at the two lowest energies, while the MR-CDFT
calculations underpredict fusion cross sections around 95 MeV.
The energy dependence of fusion cross
sections can be investigated with the fusion barrier distribution, which 
is plotted in the right panel of the figure. 
Even though the effect of anharmonicity is not large, 
one observes that 
the MR-CDFT calculation leads to a minor
improvement by considerably smearing each peak.

\begin{figure}
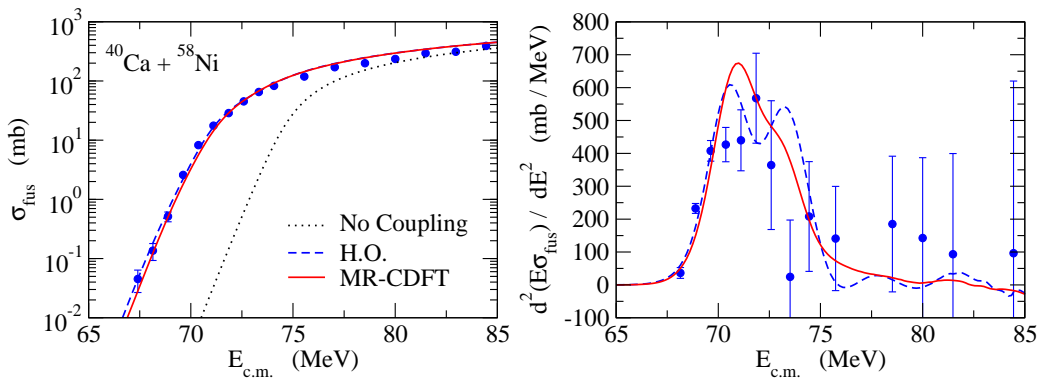

\centering
\includegraphics[scale=0.45,clip]{fig6a}
\includegraphics[scale=0.45,clip]{fig6b}
\caption{
Same as Fig. 5, but for
the $^{40}$Ca+$^{58}$Ni system.
The experimental data are taken from Ref. \cite{Bourgin14}. 
}
\end{figure}

We carry out similar calculations also for 
the $^{40}$Ca + $^{58}$Ni system \cite{HY15} (see Fig. 6). 
The effect of channel coupling is much smaller than in the 
$^{58}$Ni + $^{58}$Ni system due to a smaller charge product, $Z_PZ_T$. 
Nevertheless, we again observe that 
the anharmonicity effect in $^{58}$Ni smears
the fusion barrier distribution, leading to a somewhat 
better agreement with the
experimental fusion barrier distribution
as compared to the results in the harmonic
oscillator limit.

\section{Summary}

The research 
field of heavy-ion subbarrier fusion reactions started in the late '70s,
when a large enhancement of fusion cross sections
was experimentally discovered
with respect to the
prediction of a simple potential model.
The Wong formula
has been widely used to estimate
fusion cross sections for a given single-channel potential, 
providing reference cross sections in the absence of channel couplings 
to discuss 
the subbarrier
enhancement of cross sections. 
In this contribution, we have first extended the Wong formula by including the
energy dependence of the parameters entering the formula, that is, the
barrier height, the barrier position, and the barrier curvature. Evaluating
these parameters for the grazing angular momentum at each
energy, rather than at $l=0$, we have shown that the energy-dependent
version of Wong's formula reproduces the exact result well, even for light
systems. We have also derived a compact analytic 
formula for the oscillatory 
part of fusion cross sections, which originate from the discrete 
nature of angular momentum. 
These oscillatory parts become important in light symmetric systems, 
for which 
the symmetrization of the system amplifies the oscillations. 

We have then proposed a semi-microscopic approach to heavy-ion subbarrier 
fusion reactions. 
The basic idea of this approach is to combine 
the state-of-the-art nuclear structure calculations with 
coupled-channels calculations. For this purpose, 
we have used 
a multi-reference 
density functional theory (MR-DFT) 
based on the beyond-mean-field approach. 
The MR-DFT provides transition strengths 
among collective states without resorting to the harmonic oscillator model. 
We have applied this approach to 
the $^{58}$Ni+$^{58}$Ni and $^{40}$Ca+$^{58}$Ni fusion 
reactions, and have 
found that the anharmonicities smear the fusion barrier distributions, 
somewhat improving the agreement with the experimental data. 
It would be an interesting future problem to extend this treatment 
to heavy-ion elastic and inelastic scattering, for which 
the double-folding approach is applicable and thus 
a fully microscopic approach can be developed 
using the multi-reference density functional theory. 
That approach would also be useful in applications 
to nuclear data as well as the problem of 
nuclear transmutation \cite{Sukhovitskii00}. 

\section*{Acknowledgements}
This work was partially supported by the National Natural 
Science Foundation of China under 
Grant Nos. 11305134,  11105111, and the Fundamental Research 
Funds for the Central 
University (XDJK2013C028).

\end{document}